\documentclass[11pt]{article}

\usepackage[english]{babel}
\usepackage[pdftex]{graphicx}

\usepackage{amsmath,amsfonts,amssymb,mathrsfs}
\usepackage{hyperref}

\newtheorem{lemma}{Lemma}
\newtheorem{theorem}{Theorem}
\newtheorem{remark}{Remark}

\newcommand{\la}{\lambda}
\newcommand{\ep}{f}
\newcommand{\dl}{\delta}
\newcommand{\Tr}{{\rm Tr}}

\newcommand{\rme}{{\rm e}}
\newcommand{\rmi}{{\rm i}}
\newcommand{\rmf}{{\rm f}}

\newcommand{\uu}{{\cal M}_2}
\newcommand{\F}{\mathscr{F}}
\newcommand{\J}{{\cal F}}

\begin{document}
\title{COHERENT CONTROL OF A QUBIT\\ IS TRAP-FREE}

\author{Alexander~N.~Pechen\thanks{\href{mailto:pechen@mi.ras.ru}{
pechen@mi.ras.ru};
\href{http://www.mathnet.ru/eng/person17991}{www.mathnet.ru/eng/person17991}}
\, and Nikolay~B.~Il'in\thanks{\href{mailto:ilyn@mi.ras.ru}{ilyn@mi.ras.ru}}}
\date{\it{Steklov Mathematical Institute of Russian Academy of Sciences,\\
Gubkina str., 8, Moscow 119991, Russia.}}

\maketitle

\begin{abstract}
There is a strong interest in optimal manipulating of quantum systems by
external controls. Traps are controls which are optimal only locally but
not globally. If they exist, they can be serious obstacles to the search of
globally optimal controls in numerical and laboratory experiments, and for
this reason the analysis of traps attracts considerable attention. In this
paper we prove that for a wide range of control problems for two-level
quantum systems all locally optimal controls are also globally optimal.
Hence we conclude that two-level systems in general are trap-free. In
particular, manipulating qubits---two-level quantum systems forming a basic
building block for quantum computation---is free of traps for fundamental
problems such as the state preparation and gate generation.
\end{abstract}

\section{Introduction}

Manipulation of single quantum systems is an important branch of modern
science with applications ranging from laser-driven population transfer in
atomic systems and laser-assisted control of chemical reactions to quantum
technologies and quantum information
\cite{AlessandroBook,BrumerBook,FradkovBook,TannorBook,LetokhovBook,Brif2012}.
The 2012 Nobel Prize in Physics was awarded to Serge Harosche and David
Wineland ``for ground-breaking experimental methods that enable measuring
and manipulation of individual quantum systems'' \cite{Nobel2012}.

A fundamental issue is to control qubits, that is, two-state quantum
systems which serve as a basic building block for quantum computation and
quantum information
processing \cite{Lidar2008,Shahmoon2009,DeGreve2011,Ospelkaus2011,%
Langford2011,Blatt2012,Bocharov2012,Pierre2012,Thomas2012}.
Physical implementation of qubits includes nuclear spins addressed through
nuclear magnetic resonance, electrons in a double quantum dot controlled by
small voltages applied to the leads, holes in quantum dots controlled by
optical pulses \cite{DeGreve2011}, charge states of nanofabricated
superconducting electrodes coupled through Josephson junctions, ions in traps
\cite{Ospelkaus2011}, polarization or spatial modes of a single photon
manipulated using optical elements \cite{Langford2011}, etc. In any physical
implementation, the qubit interacts with the environment, which causes its
dynamics to be non-unitary and decreases the performance of control
operations. The simplest way to avoid the influence of the environment is to
perform fast control operations such that their duration~$T$ is significantly
smaller than the decoherence time. If this is impossible, a promising method
of dynamical decoupling \cite{Viola1999} can be used to minimize the
influence of the environment. This method has recently been experimentally
tested for the Hadamard, NOT, and $U_{\pi/8}$ gates for the gate time~$T$
exceeding the decoherence time by the order of magnitude \cite{Souza2012}.

Any physical implementation of the qubit requires the ability to optimally
prepare in a controlled manner arbitrary superpositions of the two qubit
basis states and produce arbitrary single-qubit quantum circuits. Finding
controls which optimally achieve these goals is crucial for laboratory
implementation of various quantum computing schemes \cite{Bocharov2012}.
Often the search for optimal controls is performed using numerical methods
(see, e.g., \cite{Pierre2011,Maday}) including the gradient methods (see
\cite{Glaser2010}).

\textit{Traps} are controls which are optimal only locally but not
globally. Arbitrary small variations of a trapping control do not increase
the performance of the target (e.g., a circuit operation), but globally
their outcomes can be far from good. Locally, traps look optimal, and if
they exist, they can be serious obstacles to finding desired globally
optimal controls and can significantly slow down or even completely prevent
finding such solutions in numerical and laboratory experiments. For this
reason the analysis of traps has recently attracted much
attention \cite{NOTRAPS1,NOTRAPS5,Pechen2010,PechenTannor2011,%
Moore2011,PechenTannorReply,Pechen2012,Schirmer2012,PechenTannor2012}.
Despite of these extensive studies, the absence of traps has been proved
only for the two-level Landau--Zener system \cite{Pechen2012} and for the
control of the transmission coefficient of a quantum particle passing
through a potential barrier \cite{CJC2013}. Moreover, trapping behavior has
been revealed for three-level and multi-level quantum systems
\cite{PechenTannor2011,Schirmer2012}.

The present paper contributes significantly to the field by showing that the
control of general two-level systems is completely free of traps for many
fundamental problems including those of optimal state preparation and single
qubit gate generation.

In this paper we assume that the environmental influence can be avoided so
that the Schr\"odinger equation provides a reasonable approximation for the
qubit evolution. We assume that the system is controllable so that available
controls are sufficient to produce any unitary evolution. As was shown
numerically and theoretically for the Landau--Zener system, these assumptions
can be significantly relaxed while still keeping the trap-free behavior
\cite{Pechen2012}. We also consider manipulating a single qubit. Important
problems involving control of multi-qubit dynamics, as necessary, for
example, for producing entangled states or a C-NOT gate, are beyond the scope
of this work.

\section{Formulation}
We consider coherent control of a two-level quantum system which evolves
under the action of
coherent control
$\ep(t)\in{\cal U}=L^1([0,T];\mathbb R)$ ($T>0$ is some final time) according
to the Schr\"odinger equation
\[
i\frac{dU^\ep_t}{dt}=(H_0+V\ep(t))U^\ep_t,\qquad U^\ep_{t=0}=\mathbb I
\]
Here free and interaction Hamiltonians $H_0,V\in\mathbb C^{2\times 2}$ are
two-by-two Hermitian matrices. Evolution is unitary, $U^\ep_t\in U(2)$.
The components of the matrix~$U^f_t$ belong to the space of absolutely
continuous functions on the interval~$[0,T]$, $U^f_t\in \mathrm{AC}[0,T]$.

Many important quantum control problems are terminal-time control problems,
where the goal is to maximize an objective at a specific final time~$T$.
Such objectives have the form
\[
\F(\ep)=\J(U^\ep_T)
\]
where $\J:U(2)\to\mathbb R$ is a function on the unitary group. For
definiteness, we consider maximization of the objective as the control goal,
$\F(\ep)\to\max$. The function $\J$ is assumed to be phase invariant, that
is $\J(U\rme^{i\phi})=\J(U)$ for
any $\phi$, to reflect physical equivalence of states which differ only by a
phase factor. Thus without loss of generality, we can naturally identify any
$U^\ep_T\in U(2)$ with an element of $SU(2)$ and introduce the map $\Phi:{\cal
U}\to SU(2)$ defined as $\Phi(\ep)=U^\ep_T/\sqrt{\det{U^\ep_T}}$. It is
important to emphasize that the
objective is a functional of
the control whereas $\J$ is a function of a unitary matrix.

The graph of the objective
functional $\F(\ep)$ is the \textit{dynamic control landscape}.
The graph of the function $\J(U)$ is the \textit{kinematic control landscape}.
Control $\ep$ is a \textit{trap} if $\ep$ is a local but not a global maximum of
$\F$. Control $\ep$ is a \textit{second-order trap} if $\ep$ is a critical
point, that is $\dl\F/\dl\ep=0$, Hessian of $\F$ at $\ep$ is
negative semidefinite, that is $\dl^2\F/{\dl\ep^2}\le 0$, and $\ep$ is not a
global maximum of $\F$. Control $\ep$ is \textit{regular} (or
\textit{non-degenerate}) if the differential
$D_\ep \Phi$ of the map has maximal rank. The goal of the
analysis of the control landscape is to find all traps of the objective
functional $\F(\ep)$ or to prove that there are no traps.

Among major requirements for any implementation of the
qubit are the ability to
optimally prepare arbitrary qubit states and produce arbitrary unitary
evolutions representing single-qubit quantum gates. To achieve these goals, one
has to act on the qubit with an external control $\ep(t)$, e.g. shaped laser
pulse,
small voltage, etc., which maximizes a desired objective outcome. The objective
for steering the system from the initial state
$|\rmi\rangle$ into a desired final state $|\rmf\rangle$ at time $T$ is the
transition probability
\[
 \F_{\rmi\to\rmf}(\ep)=P_{\rmi\to\rmf}=\bigl|\bigl<\rmf\bigl|U^f_T\bigr|\rmi\bigr>\bigr|^2
\]
where $U^f_T$ is the evolution operator of the system at time~$T$ induced by
the control $f$. This objective is maximized by any control $f(t)$ such that
$U^f_T|\rmi\rangle=\rme^{i\varphi}|\rmf\rangle$, where $\varphi$ is an
arbitrary (generally physically meaningless) phase. The corresponding
objective maximum is $\max_f \F_{\rmi\to\rmf}(f)=1$.

The transition probability $\F_{\rmi\to\rmf}(f)$ is a particular kind of
objectives of the form
\[
\F_O=\langle O\rangle_T=\Tr\bigl[U_T^{f}\rho_0 U^{f\dagger}_T O\bigr]
  =\Tr\bigl[\rho^f_T O\bigr],
\]
where $\rho_0$ is
the initial system density matrix and $O$ is a Hermitian operator. Such
objectives describe the problem of maximizing the average value of the system
observable $O$ at time $T$. The transition probability $\F_{\rmi\to\rmf}(\ep)$
corresponds to $\rho_0=|\rmi\rangle\langle\rmi|$ and
$O=|\rmf\rangle\langle\rmf|$. The analysis of traps for $\F_O$ for a two-level
system is equivalent to the case when $O$ is a projector. Indeed, for
a two-level system any $O$ has a representation $O=\la_1 P_1+\la_2 P_2$, where
$P_1$ and $P_2$ are two
orthogonal projectors such that $P_1+P_2=\mathbb I$, and $\la_1$ and $\la_2$ are
two eigenvalues. Thus $\Tr[\rho^\ep_TO]=\la_1+(\la_2-\la_1)\Tr[\rho^\ep_T P_2]$,
and in the non-degenerate case ($\la_1\ne\la_2$) all traps of $\Tr[\rho^\ep_TO]$
coincide with traps of $\Tr[\rho^\ep_T P_2]$. The degenerate case $\la_1=\la_2$
is trivial since in this case the objective takes the constant value
$\F(\ep)=\la_1$ and traps do not exist. Therefore without loss of generality
we
can consider $O$ as a projector, $O=|\rmf\rangle\langle\rmf|$. We denote by
$\omega_0$ and $\omega_1$ two eigenvalues of $\rho_0$ and consider
non-degenerate case $\omega_0\ne\omega_1$ since the degenerate case
$\omega_0=\omega_1=0.5$ is trivial as producing a constant objective value
$\F(\ep)=\Tr\, O$.

The objective for
generating a desired unitary gate $W$ is
\[
\F_W(f)=\frac{1}{4}\bigl|\Tr\bigl(W^\dagger U_T^f\bigr)\bigr|^2.
\]
Examples for $W$ include Hadamard gate $W=\mathbb H$, phase shift
gate $W=U_\phi$, etc. This objective is maximized by any
$U_T^\ep=\rme^{i\varphi} W$, where $\varphi$ is an arbitrary phase. The
normalization factor $1/4$ is chosen to have $\max\limits_\ep \F_W=1$ and the
absolute value is used to exclude the physically meaningless overall phase of
the unitary operator.

We consider arbitrary $H_0$ and $V$ assuming only that
$[H_0,V]\ne 0$ to have non-trivial quantum control properties. In this
case $[H_0,V]$ and $V$ are linearly independent. Indeed, assume $\alpha
V+\beta [H_0,V]=0$ for some $\alpha,\beta$ such that $|\alpha|+|\beta|>0$.
Multiplying this equality by $V$ either from the
left or from the
right and taking trace gives $\alpha\Tr(V^2)=0$, that implies $\alpha=0$, since
$V^2$ for a Hermitian $V$ is positive. Then $\beta[H_0,V]=0$ which for
$[H_0,V]\ne 0$ implies $\beta=0$. This contradicts the assumption
$|\alpha|+|\beta|>0$ and therefore $[H_0,V]$ and $V$ can not be linearly
dependent.

\section{Main result}

Let $\uu:={\rm Mat}(2,\mathbb C)$ be the complex vector space of $2\times 2$
matrices. Denote $\ep_0:=-[\Tr V\Tr H_0+2\Tr(H_0V)/[(\Tr V)^2+2\Tr(V^2)]$.
The key
result for our analysis is the following lemma.
\begin{lemma}\label{lemma1}
Let $V_t=U^{\ep\dagger}_t VU_t^{\ep\vphantom{dagger}}$ and suppose that
the function $\ep$ is not equal to the constant function $\ep_0$. Under this
assumption if a linear map $L:\uu\to\mathbb R$ satisfies $L(\mathbb I)=L(V_t)=0$
for all $t\in[0,T]$, then $L\equiv 0$.
\end{lemma}
{\bf Proof.} To prove the lemma, consider
the function $l(t):=L(V_t)$. The
equality
$L(V_t)=0$ means $l(t)\equiv 0$. Therefore, in particular,
$l(t)=l'(t)=l''(t)=0$, that implies
\begin{eqnarray}
 0 &=& L(U^\dagger_t V U^{\vphantom{\dagger}}_t)\label{eq3:4a}\\
 0 &=& L(U^\dagger_t [H_0,V] U^{\vphantom{\dagger}}_t)\label{eq3:4b}\\
 0 &=& L(U^\dagger_t ([H_0,[H_0,V]]+\ep(t)[V,[H_0,V]])
U^{\vphantom{\dagger}}_t)\label{eq3:4c}
\end{eqnarray}
We now show that if function $\ep$ is not equal to the function $\ep_0$ then
there
exists $t$ such that the matrices $\mathbb I$, $V$, $[H_0,V]$, and
$E_t=[H_0,[H_0,V]]+\ep(t)[V,[H_0,V]]$ are
linearly independent. Indeed, suppose that for all $t$
\begin{equation}\label{eq3:1}
 \alpha_t\mathbb I+\beta_t V+\gamma_t [H_0,V]+\delta_t E_t=0
\end{equation}
where complex numbers $\alpha_t$, $\beta_t$, $\gamma_t$, and $\delta_t$
satisfy
\begin{equation}\label{eq3:2}
|\alpha_t|+|\beta_t|+|\gamma_t|+|\delta_t|>0
\end{equation}
Multiplying this
equality
either by $V$ or by $H_0$ from the left and taking
trace, together with simply taking trace of Eq.~(\ref{eq3:1}), gives the system
of
equations
\begin{eqnarray*}
0&=& \alpha_t\Tr V+\beta_t\Tr V^2-\delta_t\Tr([H_0,V])^2\\
0&=& \alpha_t\Tr H+\beta_t\Tr(H_0V)+\delta_t\ep(t)\Tr([H_0,V])^2\\
0&=& 2\alpha_t+\beta_t\Tr V
\end{eqnarray*}
This system is compatible only if $\ep(t)=\ep_0$ (recall that $[H_0,V]$ is
anti-Hermitian and $[H_0,V]\ne 0$; hence $\Tr([H_0,V])^2\ne 0$). If $\ep(t)\ne
\ep_0$ for some $t$, then this system has only a trivial solution and the assumption of
linear dependence~(\ref{eq3:1}) with the requirement~(\ref{eq3:2}) leads
to contradiction. Therefore for any $t$
such that $\ep(t)\ne\ep_0$ the matrices $\mathbb I$, $V$, $[H_0,V]$ and $E_t$
are linearly independent $2\times 2$ matrices. Their unitary
evolutions $\mathbb
I$, $U^\dagger_tVU^{\vphantom{\dagger}}_t$,
$U^\dagger_t[H_0,V]U^{\vphantom{\dagger}}_t$ and
$U^\dagger_tE^{\vphantom{\dagger}}_tU^{\vphantom{\dagger}}_t$ are also linearly
independent
$2\times 2$ matrices. They form a basis of $\uu$ and hence the
equations~(\ref{eq3:4a})--(\ref{eq3:4c})
together with the assumption $L(\mathbb I)=0$ imply that $L(A)=0$ for any
$A\in\uu$. This proves the lemma.

\begin{remark}\label{r1}
The exceptional control value $\ep_0$ in the common case of
traceless interaction $\Tr V=0$ takes a simpler form
$\ep_0=-\Tr(H_0V)/\Tr(V^2)$. If, in addition, all diagonal elements of $V$ are
zero in the basis of the free Hamiltonian $H_0$ (the most common case), then
$\ep_0=0$.
\end{remark}

The exceptional control $\ep_0$ is not a trap if $T$ is sufficiently large and
$\Tr V=0$, as
stated in the following lemma.  Note that time should be large enough also
to ensure controllability of
the system.
\begin{lemma}\label{lemma2}
Let $\Tr V=0$ and $T\ge\pi/(\|H_0-\frac{1}{2}\Tr
H_0+\ep_0V\|)$, where $\|\cdot\|$ is the matrix spectral
norm. If
$\J(U)$ has no traps on $U(2)$, then the control $\ep(t)=\ep_0$ is not a trap
for $\F(\ep):=\J(U^\ep_T)$.
\end{lemma}
\noindent{\bf Proof.} The evolution of the system under the action of the
control
$\ep(t)=\ep_0+\delta\ep(t)$, where $\delta\ep$ is a small variation, is governed
by the Schr\"odinger equation
\begin{equation}\label{A1}
i\dot U^{\dl\ep}_t=(H'_0+\dl\ep(t)V)U^{\dl\ep}_t
\end{equation}
where $H'_0=H_0+\ep_0V$. The modified free Hamiltonian can be written as
$H'_0=\frac{1}{2}\Tr(H'_0)\mathbb I+H''_0$, where $H''_0$ is traceless.
The first term is proportional to the identity matrix and can be neglected. The
second term in the suitable basis can be written as $H''_0=\omega_0\sigma_z$,
$\omega_0>0$ and by suitably rescaling time we can set
$\omega_0=1$. Thus,
instead of the evolution equation~(\ref{A1}) we can consider the equivalent
equation
\begin{equation}\label{A2}
 i\dot U^{\dl\ep}_t=(\sigma_z+\dl\ep(t)V)U^{\dl\ep}_t
\end{equation}
Checking if $\ep_0$ is not a trap for eq.~(\ref{A1}) is equivalent to
checking if $\ep(t)=0$ is not a trap for eq.~(\ref{A2}).

The interaction can be written as
$V=v_x\sigma_x+v_y\sigma_y+v_z\sigma_z+v_0\mathbb I$. We consider the
non-trivial case $v=\sqrt{v_x^2+v_y^2}\ne 0$.
The evolution operator produced by $\dl\ep(t)=0$ has the form
$U^0_t=\rme^{-it\sigma_z}$. Introducing the angle $\phi=\arctan(v_y/v_x)$, we can
write $V_t:=V^0_t:=U^{0\dagger}_t V
U^0_t=v\cos(2t-\phi)\sigma_x-
v\sin(2t-\phi)\sigma_y+v_z\sigma_z+v_0$. This gives
for the gradient of the objective
\[
\nabla
\F_{\ep_0}(t)=v\cos(2t-\phi)L(\sigma_x)-v\sin(2t-\phi)L(\sigma_y)+v_zL(\sigma_z)
\]

Suppose $v_z\ne 0$ or $L(\sigma_z)=0$. If $\ep_0$ is a critical point, then
the gradient $\nabla \F_{\ep_0}(t)=0$
for any $t\in[0,T]$ and, therefore
$L(\sigma_x)=L(\sigma_y)=0$. In addition, $L(\mathbb I)=0$ for any phase-invariant objective
and hence $L\equiv 0$ on
$\uu$. Then similarly to the proof of the Theorem~1 we conclude that
$\ep=\ep_0$ is not a trap (it can be either a global maximum or a global
minimum).

Now consider the case $v_z=0$ and $L(\sigma_z)\ne 0$. For this case we assume in
addition
that the interaction is traceless, that is $v_0=\Tr V=0$. The evolution
operator
produced
by a small variation of the control
$\dl\ep$ can be represented as
$U^{\dl\ep}_T=U^0_T\widetilde{U}^{\vphantom{0}}_T$, where $U^0_T=\rme^{-i
T\sigma_z}$ and $\widetilde{U}_T$ satisfies
\[
\dot{\widetilde{U}}{}^{\dl\ep}_t=-i\dl\ep(t)V^0_t\widetilde{U}{}^{\dl\ep}_t,\qquad
\widetilde{U}{}^{\dl\ep}_0=\mathbb I
\]
The operator $\widetilde{U}_T$ can be computed up to the second order in
$\dl\ep$ as
\begin{eqnarray*}
 \widetilde{U}{}^{\dl\ep}_T&=&\mathbb I+A_1+A_2+o(\|\dl\ep\|^2)\\
 && A_1=-i\int_0^T dt\dl\ep(t)V^0_t,\\
 && A_2=-\int_0^Tdt_1\int_0^{t_1}dt_2\dl\ep(t_1)\dl\ep(t_2)V^0_{t_1}V^0_{t_2}
\end{eqnarray*}
We choose $\dl\ep_1$ and $\dl\ep_2$ such that $A_1=0$, that is,
\begin{equation}\label{A3}
\int\limits_0^T dt\dl\ep_k(t)\cos 2t=\int\limits_0^T dt\dl\ep_k(t)\sin 2t
=0\quad (k=1,2)
\end{equation}
For such $\dl\ep_k$ noting that
$V^0_{t_1}V^0_{t_2}=v^2[\cos2(t_1-t_2)+i\sigma_z\sin2(t_1-t_2)]$, we get
$A_2=-i I(\dl\ep_{k})\sigma_z$, where
\[
I(\dl\ep)=v^2\int_0^Tdt_1\int_0^{t_1}dt_2\dl\ep(t_1)\dl\ep(t_2)\sin2(t_1-t_2).
\]
Then, up to the second order in $\dl\ep$ we have
\begin{eqnarray*}
\F(\dl\ep)&=&\J(U^0_T(\mathbb
I+A_2+\dots))\\
&=&\J(U^0_T)+\Tr\left(\frac{\dl\J}{\dl
U}\biggl|_{U=U^0_T} U^0_T A_2\right)+o(\|\dl\ep\|^2)\\
&=&\J(U^0_T)+I(\dl\ep)L(\sigma_z)+o(\|\dl\ep\|^2)
\end{eqnarray*}

Now suppose $T\ge\pi$ in the rescaled time frame (that corresponds to
$T\ge\pi/(\|H_0-\frac{1}{2}\Tr H_0+\ep_0V\|)$ in the original time frame). We will show the
existence of variations
$\dl\ep_1$ and $\dl\ep_2$ which satisfy Eq.~(\ref{A3}) and
produce $I(\dl\ep_1)$ and $I(\dl\ep_2)$ with
opposite signs. An example is $\dl\ep_1(t)=\chi_{[0,\pi]}(t)$ and
$\dl\ep_2(t)=\cos
(4t)\chi_{[0,\pi]}(t)$, where $\chi_{[0,\pi]}(t)$ is the characteristic
function of the interval $[0,\pi]$. For these variations $I(\dl\ep_1)=\pi v^2/2$
and
$I(\dl\ep_2)=-\pi v^2/12$. Therefore for $L(\sigma_z)\ne 0$ there exist
directions at $\ep(t)=0$ in which the objective increases and
directions in which it decreases. This means that $\ep(t)=0$ for Eq.~(\ref{A2})
(and thus $\ep(t)=\ep_0$ for Eq.~(\ref{A1})) is neither a local maximum nor a
local minimum, and hence is not a trap. This proves the lemma.

Our main result is the following theorem.

\begin{theorem}\label{th1}
Suppose the only extrema of the kinematic landscape $\J(U)$ are global maxima
and global minima. If $\Tr V=0$ and $T\ge\pi/(\|H_0-\frac{1}{2}\Tr H_0+\ep_0V\|)$, then the
only extrema of the dynamic landscape $\F(\ep)$ are global maxima and global
minima.
\end{theorem}
\noindent{\bf Proof.} Consider first the case $\ep(t)\ne\ep_0$. The variation
of $U_T^f$ has the
form $\delta U_T^f/\delta\ep(t)=-i U_T^f V_t^f$, where $V^f_t=U_t^{f\dagger}V U_t^f$. By the chain rule,
\[
\frac{\delta \F}{\delta\ep(t)}=\Tr\left(\frac{\delta\J}{\delta
U}\biggl|_{U=U_T^f}\frac{\delta U_T^f}{\delta \ep(t)} \right) =
-i\Tr\left(\frac{\delta \J}{\delta U} U_T^f V_t^f\right)=:L(V_t^f)
\]
Denoting $X=-i(\delta \J/\delta U) U_T$, we get $L(A)=\Tr(XA)$. The
assumption $\J(U\rme^{i\phi})=\J(U)$ for any $\phi$ implies that $L(\mathbb
I)=0$. Indeed, then
\[
0=\frac{\partial
\J(U_T^f\rme^{i\phi})}{\partial\phi}\biggl|_{\phi=0}=i\Tr\left(\frac{\delta
\J}{\delta U} U\biggl|_{U=U_T^f}\right)=-L(\mathbb I)
\]
If $\ep(t)$ is a critical control, then also $L(V_t)=0$ and the Lemma implies
$L\equiv 0$. Taking $A=X^\dagger$, we get $L(X^\dagger)=\Tr(XX^\dagger)=0$ and
therefore $X=0$. Since $U^\ep_T$ is unitary, that implies $\delta\J/\delta U=0$,
i.e. $\ep$ is an extrema of the functional $\F(\ep)$ if and only if $U=U^\ep_T$
is an extrema of the function $\J$. Hence if the only extrema of $\J$ are global
maxima and global minima, then the same is true for $\F(\ep)$ apart possibly of
the exceptional control $\ep=\ep_0$. The control $\ep(t)=\ep_0$ requires a
separate analysis and is shown to be not a
trap in Lemma~\ref{lemma2}. This completes the proof.

\begin{remark}\label{r2}
The statement of Theorem~\ref{th1} is non-trivial and is a special property of
two-level
systems. In general, the
trap-free property of $\J(U)$ might not imply the trap-free property
of
$\F(\ep)$ as was shown for various $n$-level systems with $n\ge
3$~\cite{PechenTannor2011,Schirmer2012}.
\end{remark}

\begin{remark}
The statement of Lemma \ref{lemma1} means that the map $f\to U^f_{T}$ has the
maximal rank at each point~$U^f_{T}$ of the unitary group $\mathrm{SU}(2)$
because the gradient $\nabla_{f} U^f_{T}$ is surjective on the tangent bundle
of~$\mathrm{SU}(2)$. In this case, as follows from \cite[Theorem~1]{new}, the
critical points of the kinematic landscape are in bijective correspondence
with the critical points of the dynamic landscape. Thus, if the kinematic
landscape has not only global maxima and global minima but also saddle
points, then the statement of Theorem~\ref{th1} about the absence of traps
remains valid.
\end{remark}

While this theorem can be used to prove the absence of traps for objectives
$\F_O$ and $\F_W$, below we treat these important cases independently.

\begin{theorem}\label{th2}
Let $\Tr V=0$ and $T\ge\pi/(\|H_0-\frac{1}{2}\Tr H_0+\ep_0V\|)$. Then the only extrema of
$\F_O(\ep)$ (hence also of $\F_{\rmi\to\rmf}(\ep)$ as well) are
global maxima and global minima.
\end{theorem}
{\bf Proof.} Consider first the case $\ep\ne\ep_0$. The gradient of the objective
$\F_O$ is $\nabla \F_O(t)=L_O(V_t)$, where the map $L_O:\uu\to\mathbb R$ is
defined by
\[
L_O(A)=-i\Tr([\rho_0, O_T]A)
\]
with $O_T=U^{f\dagger}_T O U_T^{f}$. At any critical control,
$\nabla \F_O=0$ and hence
$L_O(V_t)=0$. Clearly, $L_O(\mathbb I)=0$. Then the Lemma implies
that $L_O\equiv 0$. In this proof, we denote by
$|\rmf\rangle$ vector such that $O_T=|\rmf\rangle\langle\rmf|$, and
denote by $|\rmf_\bot\rangle$ vector orthogonal to $|\rmf\rangle$. Now take
the operators
$A=|\rmf_\bot\rangle\langle\rmf|+|\rmf\rangle\langle\rmf_\bot|$ and
$A'=i(|\rmf_\bot\rangle\langle\rmf|-|\rmf\rangle\langle\rmf_\bot|)$. The
equalities
$L_O(A)=0$ and $L_O(A')=0$ imply ${\rm Im}\langle\rmf_\bot|\rho_0|\rmf\rangle=0$ and
${\rm Re}\langle\rmf_\bot|\rho_0|\rmf\rangle=0$, respectively. Hence
$\langle\rmf_\bot|\rho_0|\rmf\rangle=0$ and therefore
$|\rmf\rangle$ is an eigenstate of $\rho_0$. Its only possible eigenvalues are
$\omega_0$ and $\omega_1$ that correspond to the global minimum
($\F_O^{\min}=\langle\rmf|\rho_0|\rmf\rangle=\omega_0$) and the global maximum
($\F_O^{\max}=\langle\rmf|\rho_0|\rmf\rangle=\omega_1$) of the
objective, respectively. These are the only allowed critical points except of
$\ep(t)\equiv \ep_0$. The proof for the exceptional case $\ep_0$ follows from
Lemma~\ref{lemma2}.

\begin{theorem}\label{th3}
Let $\Tr V=0$ and $T\ge\pi/(\|H_0-\frac{1}{2}\Tr H_0+\ep_0V\|)$. Then the only extrema of
$\F_W(\ep)$ are global maxima and global minima.
\end{theorem}
{\bf Proof.} First we consider the case $\ep\ne\ep_0$. The gradient of the objective
$\F_W$ has the form $\nabla
\F_W(t)=L_W(V_t)$, where the map $L_W:\uu\to\mathbb
R$ is defined by
\[
L_W(A)=\frac{1}{2}\left[\Im\Tr Y\cdot\Re\Tr(YA)-\Re\Tr Y\cdot\Im\Tr(YA)\right]
\]
Here $Y=W^\dagger U^\ep_T$ ($Y$ is unitary). Clearly,
$L_W(\mathbb I)=0$. At any critical control $\nabla \F_W=0$ and hence
Lemma~\ref{lemma1} implies that $L_W\equiv 0$. We consider the
operators $A=Y+Y^\dagger$ and $A'=i(Y-Y^\dagger)$ and denote $\Tr
Y=y_\Re+i y_\Im$ and $\Tr Y^2=z_\Re+i z_\Im$, where $y_\Re$ and $y_\Im$ are real
and
imaginary parts of $\Tr Y$, $z_\Re$ and $z_\Im$ are real and
imaginary parts of $\Tr Y^2$. The equalities $L_W(A)=L_W(A')=0$ become
\begin{eqnarray*}
z_\Im y_\Re-(z_\Re+2)y_\Im=0\\
(z_\Re-2)y_\Re+z_\Im y_\Im=0
\end{eqnarray*}
The solution $y_\Re=y_\Im=0$ corresponds to the global minimum of the objective
($\F^{\rm min}_W=0$). The compatibility of the system for other solutions
requires
$z^2_\Im+z_\Re^2\equiv |\Tr Y^2|^2=4$. Since $\F=(1/4)|\Tr Y|^2=(1/4)|\Tr
Y^2|^2$, that implies that these solutions correspond to the global maximum
$\F^{\rm
max}_W=1$ and no other solutions exist. The proof for the
exceptional case $\ep_0$ follows from Lemma~\ref{lemma2}.

\section*{Acknowledgements}
This work was supported by the Russian Foundation for Basic Research, project
no.~14-01-31115.

\end{document}